\documentclass[pdflatex,sn-mathphys-num]{sn-jnl}

\AtBeginDocument{%
  \fontsize{11bp}{13bp}\selectfont
  \def\listfont{\fontsize{11bp}{13bp}\selectfont}%
}

\usepackage{graphicx}
\usepackage{multirow}
\usepackage{amsmath,amssymb,amsfonts}
\usepackage{amsthm}
\usepackage{mathrsfs}
\usepackage[title]{appendix}
\usepackage{xcolor}
\usepackage{textcomp}
\usepackage{manyfoot}
\usepackage{booktabs}
\usepackage{algorithm}
\usepackage{algorithmicx}
\usepackage{algpseudocode}
\usepackage{listings}
\theoremstyle{thmstyleone}

\theoremstyle{thmstyletwo}

\theoremstyle{thmstylethree}

\begin{document}

\title[Article Title]{$ $ $ $ $ $ $ $ $ $ $ $ $ $ $ $ $ $ $ $ $ $ $ $ $ $ $ $ $ $ $ $ $ $ $ $ $ $ The Data Heat Island Effect: \\Quantifying the Impact of AI Data Centers in a Warming World}

\author*[1,2]{\fnm{Andrea} \sur{Marinoni}}\email{am2920@cam.ac.uk}


\author[3]{\fnm{Erik} \sur{Cambria}}\email{cambria@ntu.edu.sg}


\author[3]{\fnm{Weisi} \sur{Lin}}\email{wslin@ntu.edu.sg}

\author[5]{\fnm{Mauro} \sur{Dalla Mura}}\email{mauro.dalla-mura@gipsa-lab.grenoble-inp.fr} 

\author[6]{\fnm{Jocelyn} \sur{Chanussot}}\email{jocelyn.chanussot@grenoble-inp.fr} 

\author[7]{\fnm{Edoardo} \sur{Ragusa}}\email{edoardo.ragusa@unige.it}

\author[8]{\fnm{Chi Yan} \sur{Tso}}\email{chiytso@cityu.edu.hk} 

\author[9]{\fnm{Yihao} \sur{Zhu}}\email{yihaozhu2@cityu.edu.hk} 

\author[8]{\fnm{Benjamin} \sur{Horton}}\email{bphorton@cityu.edu.hk}


\affil*[1]{\orgname{Department of Computer Science and Technology, University of Cambridge}, \orgaddress{\street{15 JJ Thomson Avenue}, \city{Cambridge}, \postcode{CB3 0FD}, \country{UK}}}

\affil[2]{\orgname{Glitch Analytics Ltd.}, \orgaddress{\street{The Old School House, Church Lane}, \city{Oving}, \postcode{HP22 4HL}, \country{UK}}}

\affil[3]{\orgname{College of Computing and Data Science, Nanyang Technological University}, \orgaddress{\street{50 Nanyang Avenue}, \postcode{639798}, \country{Singapore}}}

\affil[4]{\orgname{Asian School of Environment, Nanyang Technological University}, \orgaddress{\street{50 Nanyang Avenue}, \postcode{639798}, \country{Singapore}}}

\affil[5]{\orgname{GIPSA-lab, Grenoble-INP UGA}, \orgaddress{\street{11 rue des Mathématiques, Grenoble Campus BP46}, \city{Saint-Martin-d'Hères Cedex}, \postcode{38402}, \country{France}}}

\affil[6]{\orgname{INRIA}, \orgaddress{\street{655 Avenue de l'Europe, CS 90051}, \city{Montbonnot Cedex}, \postcode{38334}, \country{France}}}

\affil[7]{\orgname{Dipartimento di Ingegneria Navale, Elettrica, Elettronica e delle Telecomunicazioni - DITEN, University of Genoa}, \orgaddress{\street{via alla Opera Pia 11a}, \city{Genoa}, \postcode{16145}, \country{Italy}}}

\affil[8]{\orgname{School of Energy and Environment, City University of Hong Kong}, \orgaddress{\street{Tat Chee Avenue, Kowloon}, \country{Hong Kong}}}

\affil[9]{\orgname{Department of Management, College of Business, City University of Hong Kong}, \orgaddress{\street{Tat Chee Avenue, Kowloon}, \country{Hong Kong}}}


\abstract{
The strong and continuous increase of AI-based services leads to the steady proliferation of AI data centres worldwide with the unavoidable escalation of their power consumption. 
It is unknown how this energy demand for computational purposes will impact the surrounding environment. 
Here, we focus our attention on the heat dissipation of AI hyperscalers. 
Taking advantage of land surface temperature measurements acquired by remote sensing platforms over the last decades, we are able to obtain a robust assessment of the temperature increase recorded in the areas surrounding AI data centres globally. 
We estimate that the land surface temperature increases by 2°C on average after the start of operations of an AI data centre, inducing local microclimate zones, which we call the \textit{data heat island effect}. 
We assess the impact on the communities, quantifying that more than 340 million people could be affected by this temperature increase. 
Our results show that the data heat island effect could have a remarkable influence on communities and regional welfare in the future, hence becoming part of the conversation around environmentally sustainable AI worldwide.}


\maketitle

\section{Introduction}
\label{sec:introduction}
When considering the impact of anthropogenic activities on climate change and global warming in the last few decades, the urban heat island (UHI) effect plays a key role.  
UHI results from the concentration of industrial activities and heavy use of synthetic construction material, as well as solid increase in energy consumption in densely populated urban areas ~\cite{UHI_1, UHI_2, UHI_4, UHI_5, UHI_6, UHI_7}.  
When discussing UHI effects, it is paramount to consider the impact they have on local communities and regional welfare. 
In this respect, it has been studied and demonstrated how UHI would affect healthcare, energy consumption, air quality, and water quality. 
It is therefore crucial to understand their causes~\cite{UHI_11}. 

Specifically, the main drivers for UHI are classified in terms of geometry of the anthropogenic spaces (e.g., urban canyons dictate the concentration of particulates); lack of vegetation and water bodies; generation of air pollutants and water vapour; heat retention and low albedo building~\cite{UHI_1,UHI_4,UHI_7,UHI_8,UHI_10}. 
On top of these categories, the type of human activities established and operating  dramatically influence the impact of UHI on environment and communities. 
These anthropogenic activities are characterized by their spatial concentration, functional type (e.g., residential, industrial), cooling requirements, fuel sources, and thermal efficiency~\cite{UHI_8,UHI_1, UHI_11}. 

With global data volumes growing rapidly~\cite{camsto}, data centres are expected to be one of the most power-hungry activity in the next decade~\cite{Alberta, PowerAI, PowerHungry, MSOpenAINvidia}. 
In fact, it has been estimated that in 3 to 5 years the power consumption for data processing will exceed the amount budgeted for manufacturing~\cite{McKinsey, McKinsey2, NatClimChange_Masanet}. 
As such, it is possible to expect that the impact of data centres and AI hyperscalers activities on climate might not be negligible~\cite{AIEnergyConsum, IEA, Science_masanet}, indeed being further exacerbated by the use of AI in the next decades \cite{McKinsey, McKinsey2, Bloomberg}. 

In fact, AI data centres are in vast majority relying on fossil fuel use~\cite{IEA, Science_masanet, McKinsey, Bloomberg, JLL}. 
Therefore, the steep growth of AI training and use for various applications would directly translate into high net impact on emissions. Also, the inefficiencies and nonidealities of AI hyperscalers operations would cause their emissions to rise even more under the expected projections of AI model scaling~\cite{IEA}. 

However, studying in detail the actual impact of AI hyperscalers environmental footprint would entail several uncertainties and unknowns~\cite{IEA}. 
Indeed, there are no credible indicators that can help determine and quantify the impact of existing AI applications. 
Moreover, predicting the characteristics of future AI applications, particularly their energy consumption and computing requirements, remains challenging. 
The fluctuations on forecast scenarios that could be drawn are further amplified by a lack of both consistency in methodologies and comprehensive data, even in historic estimates of environmental footprint from the ICT sector~\cite{IEA,Science_masanet, SustainableAI,UHI_9}.
\begin{figure}[h]
\centering
\includegraphics[width=1\textwidth]{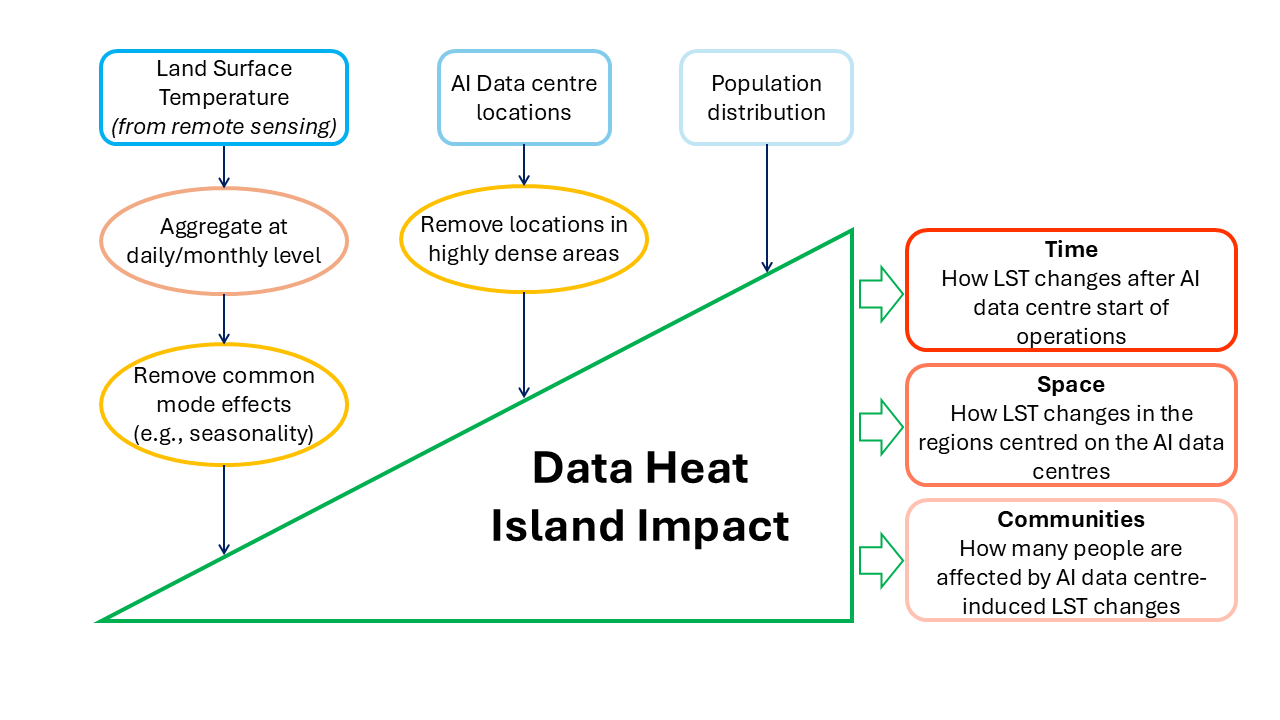}
\caption{Graphical abstract of this work: the proposed multiscale multimodal architecture for data analysis integrates records of land surface temperature trends from year 2004 to 2024, gridded population maps, and AI data centres locations worldwide, to achieve a thorough understanding of the impact of \textit{data heat islands} in time, space and over communities.}
\label{fig_workflow}
\end{figure}
In this paper, we aim to provide a novel perspective on the impact of AI hyperscalers on environment and sustainability under various climate scenarios for the next decades. 
By a multiscale multimodal analysis of records collected from various sources of information and sensors, we quantify the contribution of AI hyperscalers worldwide on land surface temperature increase (see Figure \ref{fig_workflow}). 
Specifically, we integrate land surface temperature data from remote sensing platforms with the locations of AI hyperscalers that have been established in the last twenty years in order to assess the change in atmospheric heat induced by the data centres. 
To make our study less prone to uncertainties, we harmonise, arrange and filter the data from seasonality effects, influence from various nonidealities (e.g., missing data), and influence from factors (e.g., other types of anthropogenic activities in the surrounding of AI hyperscalers) which might contaminate our study. In this way, we are able to identify the impact of AI hyperscalers on environment as a consistent gradient in temperature between the area of data centres and their surrounding regions, so much so that could form a “\textit{data heat island effect}”. 
Our analysis moves in order to address three main objectives:
\begin{itemize}
    \item Quantification of the land surface temperature increase connected to the establishment of an AI hyperscalers; 
    \item Assessment of the region of influence of this increase; 
    \item Estimation of the population affected by the temperature increase.
\end{itemize}
The paper is organised as follows. 
Section \ref{sec:methods} introduces the datasets and the methodology used to extract information on the land surface temperature increase induced by the AI data centres in  twenty years worldwide. 
Section \ref{sec:results} reports the results we achieved with respect to the aforementioned indicators. 
Section \ref{sec:way_forward} provides a few best practice guidelines for software and hardware solutions that could help to reduce the impact of the data heat island effect. 
Finally, Section \ref{sec:conclusions} draws the final remarks and conclusions.

\section{Data and methods} 
\label{sec:methods}

Our analysis of the environmental footprint of AI hyperscalers relies primarily on land surface temperature (LST) measurements (see Figure \ref{fig_workflow}). 
We used a reconstructed MODIS LST dataset (produced by NASA) acquired worldwide from 2004 to 2024 over an enhanced 500m resolution grid~\cite{MODIS_1}. 
To address data limitations (e.g., missing acquisitions, cloud cover), we aggregated the results at daily and then monthly scale, and removed seasonality effects, as well as outliers. 
Then, we focused our attention on the locations of the main AI hyperscalers that have been built in the same time interval, collecting information from resources such as \cite{DataCenter_data}. 
This database contains more than 11000 locations worldwide, of which 8472 have been detected to dwell outside of highly dense urban areas. 
We thus used the latter locations to quantify the effect of the establishment of data centres on the environment in terms of the LST gradient that could be measured on the areas surrounding each data centre. 
In fact, LST profiles in urban and densely built-up areas can be affected by various activities, e.g., manufacturing, house heating, road networks. 
Considering only the AI data centres located outside of highly dense regions allows us to provide a solid connection between the LST trends that we can measure and the presence of AI data centres in the area. 
In particular, we assess mean LST trends over time within circular regions centered on each AI hyperscaler. 
In practice, our analysis counts over 6733 data points, when considering LST trends cleared out of problematic data points and outliers for AI data centres located outside of highly dense urban areas. 
Finally, to assess the impact of LST change on population, we downscale the 100m x 100m demographic maps from Worldpop Global Project~\cite{Worldpop} over the considered to 1km resolution, so to enable a robust statistical analysis of the population affected the temperature increase induced by AI data centres. 

This study relies on the assumption that AI hyperscalers might have an impact on the LST of their locations because of the heat that they would release as a result of the high power demanding applications that they would be used for~\cite{Verne_AIheat}. 
To assess this, we first quantify the normalised temperature increase that could be observed in the months right after the start of operations of the AI hyperscalers we considered. 
To assess this impact we draw inspiration from  the quantification approach used to estimate the urban heat island effect \cite{UHI_7,UHI_8,UHI_10}.  
Specifically, we compute the monthly average LST of the area 
centred on each AI hyperscaler, and we calculate the difference between the mean monthly LST and average LST that is recorded across a period of \textit{k} months on that region. 
In other terms, let us define $T_i^r$ as the mean LST for month $i$ at a distance of $r$ km from a given data centre. 
Then, for every AI data centre location, we can write the normalised temporal temperature increase centred over the data centre location as $\Delta_i^{r=0}(k) = \Delta_i^{0}(k)$, which is defined follows: 
\begin{equation}
    \Delta_i^0(k) = T_i^0 - \frac{1}{k}\sum_{j=1}^k T^0_{i-j},
    \label{eq_DeltaT}
\end{equation}
where $T_i^0$ identifies the mean LST for month $i$ for each AI data centre location. 
For completeness, this quantity is typically computed over 10 years, which translates into setting $k=120$ in (\ref{eq_DeltaT}).

To focus our attention on the impact of AI data centres on LST increase, we centre the origin of the $i$-axis on the date of start of operations of each data centre under exam. 
Therefore, $\Delta_0(k)$ identifies the average LST increase that is measured over each AI data centre with respect to the mean of the LST that was recorded over the $k$ months before their start of operations. 

With this in mind, we assess the spatial influence of AI data centres on LST increase by focusing our attention on the LST distribution at $r$ km from the given data centres over the $k$ months prior to the start of operation of each data centre, that is:
\begin{equation}
    \Delta_0^r(k) = \overline{T}_0^r - \frac{1}{k}\sum_{j=1}^k \overline{T}^r_{0-j},
    \label{eq_DeltaT_space}
\end{equation}
\noindent where $\overline{T}_0^r$ and $\overline{T}_{-j}^r$ identify the average of the LST of all the points at $r$ km from the given AI data centre on the month of  start of operation and at $j$ months prior, respectively. In other terms, assuming 
that $R$ points are at $r$ km from the given data centre, and that $\left. T_u^r \right|_l$ identifies the LST that we estimate for the $l$-th point at $r$ km from the given data centre and $u$ months from the start of operations, we can write $\overline{T}_u^r = \sum_{l=1}^R \left. T_u^r \right|_l /R$. 

As mentioned, we compute the quantities in (\ref{eq_DeltaT}) and (\ref{eq_DeltaT_space}) for each data centre under exam. 
The following Section reports the results we achieved building up on these metrics.

\section{Results and discussion} 
\label{sec:results}
\begin{figure}[h]
\centering
\includegraphics[width=1\textwidth]{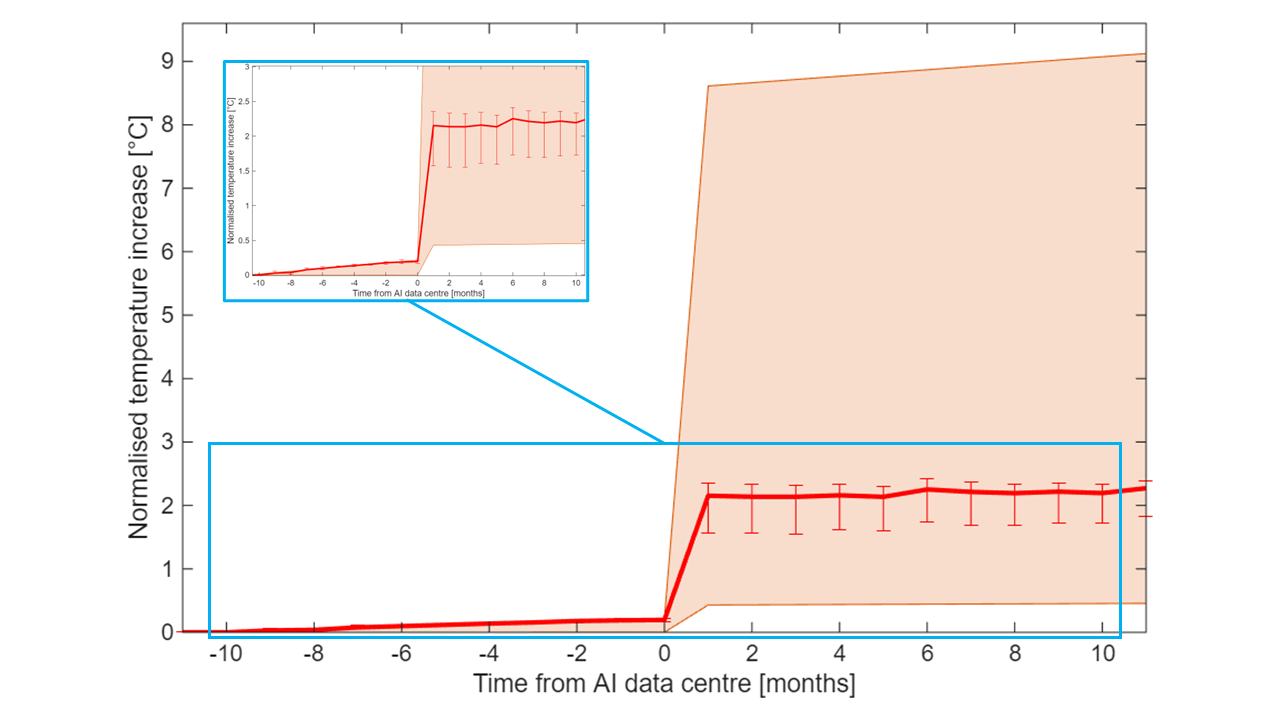}
\caption{Temperature increase through time over the AI hyperscalers locations centred around the time of start of operations ($i=0$), according to the procedure described in Section \ref{sec:results} - equation (\ref{eq_DeltaT}). 
The aggregate average of the LST difference  is shown in red solid line. 
The shaded areas show the interval between the maximum and minimum value of LST increase that has been recorded across the considered AI hyperscalers. 
The bar across the average line identifies the limit of the 95th percentile of the distribution we compute.}
\label{fig_TempIncrease}
\end{figure}

\begin{figure}[h]
\centering
\includegraphics[width=1\textwidth]{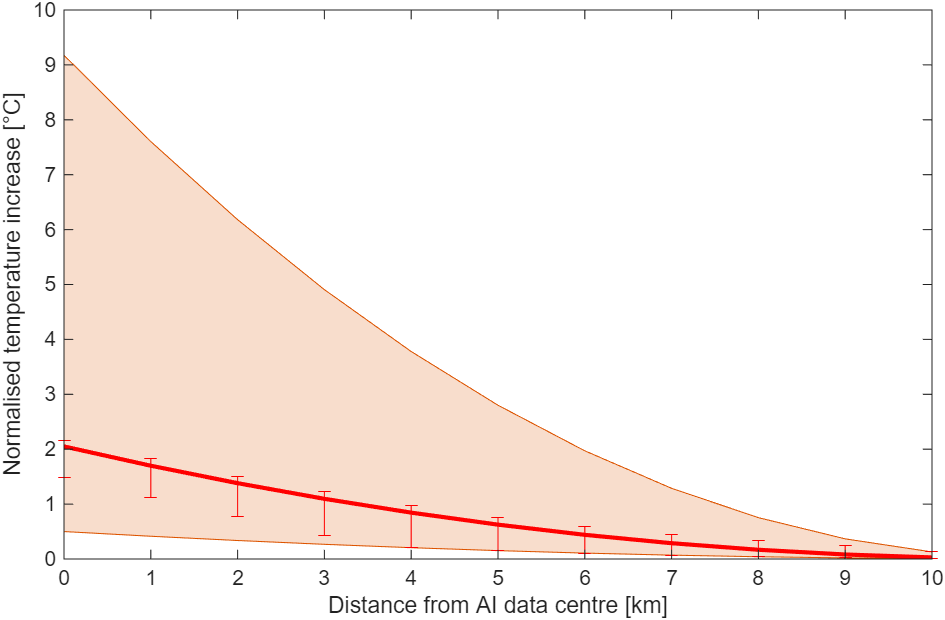}
\caption{Temperature increase through space as a function of the distance from the AI hyperscalers locations, according to the procedure described in Section \ref{sec:methods} - equation (\ref{eq_DeltaT_space}). 
The same color policy as in Figure \ref{fig_TempIncrease} applies here.}
\label{fig_TempRadius}
\end{figure}

\begin{figure}[h]
\centering
\includegraphics[width=1\textwidth]{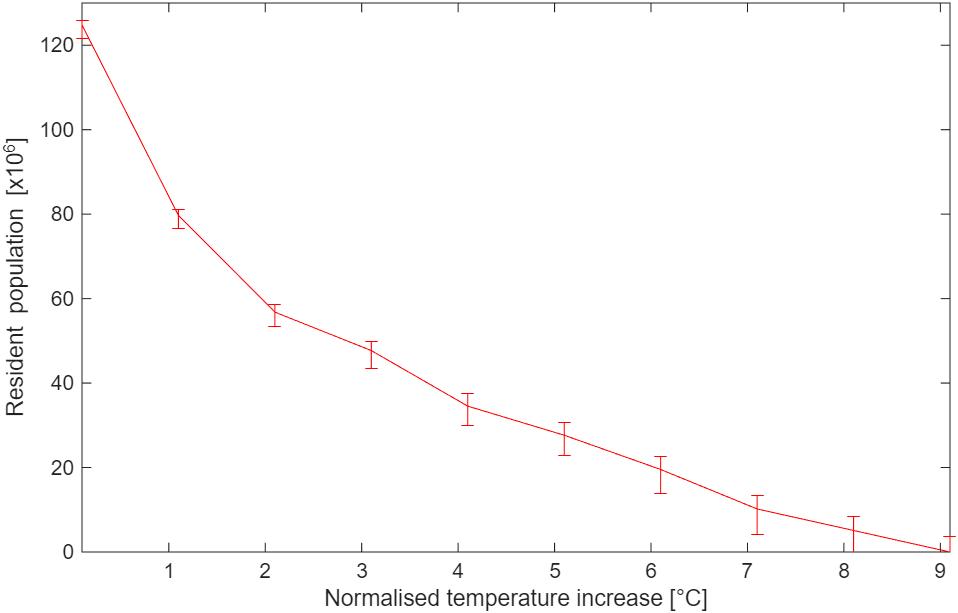}
\caption{Distribution of population 
as a function of the LST increase they are affected by
within $r=10$ km radius from the AI hyperscalers considered in this study with respect to the LST trend of the 5 years prior to the start of operations of the data centres, i.e., $\Delta_0^{10}(60)$ in (\ref{eq_DeltaT_space}) in Section \ref{sec:methods}.
}
\label{fig_TempPop}
\end{figure}

The results we obtained conducting the analysis described Section~\ref{sec:methods}  across all AI hyperscalers analysed during the 2004-2024 period are aggregated and displayed in Figure~\ref{fig_TempIncrease}. 
Specifically, we focus our attention on the LST increase induced over the AI data centres under exam with respect to the average LST recorded over those regions for the 5 years prior to each AI data centre start of operations. 
For sake of visualisation, we concentrate 10 months before the start of operations of each date centre, and the 10 months after it. 
In other words, we compute $\Delta_i^0(k)$ as in (\ref{eq_DeltaT}) with $k=5 \times 12 = 60$, and for $i \in \{-10, -9, ..., -1, 0, 1, ... 9, 10\}$.
In particular, we align the temporal analysis results over the x-axis, where the trends are centred over the time of start of operations of each data centre. 
The aggregate average of the LST difference is shown in red solid line. 
The shaded areas show the interval between the maximum and minimum value of LST increase that has been recorded across the considered AI hyperscalers. 
Finally, the bar across the average line identifies the limit of the 95th percentile of the distribution we compute. 
Figure~\ref{fig_TempIncrease} shows a clear increase of LST coinciding with the start of operations of the AI hyperscalers that have been monitored in this study. 
Indeed, the average LST increase across the data centres is 2.07°C, whilst its minimum  and maximum can be found at 0.3 °C and 9.1 °C, respectively. 
The 95th percentile of the LST increase after the AI data centres start of operations is concentrated between 1.5°C and 2.4°C. 
These results are dramatically impressive, especially considering that the typical LST increase caused by the quintessential example of compound of anthropogenic activities – the urban heat island effect – has been estimated in the 4 and 6 °C interval ~\cite{UHI_1,UHI_8,UHI_7}. 
This apparent step function emphasize the clear effect of AI hyperscalers on their surrounding areas, so much that it can match the impact of ``islands'' of higher temperatures: therefore, we call this the \textit{data heat island} effect. 
This terminology is further supported when computing monthly LST differences as in Equation~\eqref{eq_DeltaT} over various time intervals (12 months to 10 years). 
Table~\ref{tab_tempIncrease} reports these results: across all the intervals that have been considered, the LST increase over the AI hyperscaler regions over the start of operations of the data centre seems consistent.  
\begin{table}[h]
\caption{Temperature increase (in °C) as defined in Equation~\eqref{eq_DeltaT} at AI hyperscalers start of operation (i=0) as a function of $k$.}
\label{tab_tempIncrease}%
\begin{tabular}{@{}lllll@{}}
\toprule
$\Delta_0(k)$ & \textbf{k=12}  & \textbf{k=24} & \textbf{k=36} & \textbf{k=120}\\
\midrule
 average  &  2.03  & 2.05 & 2.06  & 2.12 \\
 minimum  &  0.30  & 0.31 & 0.32  & 0.37 \\
 maximum  &  9.02  & 9.09 & 9.14  & 9.24 \\
\botrule
\end{tabular}
\end{table}
The influence of AI hyperscalers apparently is not limited to the immediate proximity of their locations. 
In fact, we computed the temperature increase over wider regions circularly arranged around the data centres, following the same procedure that we previously described. Figure \ref{fig_TempRadius} displays the results of this analysis. 
Taking a look to these results, it is evident that the impact of LST increase reaches up to 10 km distance from the AI hyperscalers. 
The data heat island effect seems to reduce its intensity to 30$\%$ within 7 km around the data centres. 
In particular, an average monthly LST increase of 1 °C can be measured up to 4.5 km from the AI hyperscalers. 
This spatial extent is comparable to that observed in urban heat island effects~\cite{UHI_1,UHI_7,UHI_8}. 
The LST increase that is recorded in the area surrounding AI data centres seem consistent across various regions of the world, even if under diverse climatic conditions. 
As an example, we mention three situations that show the consistency of our study. 
In fact, in the following case studies, our LST increase assessment fit some anomalous trends of LST increase that have been recorded in the last decades, so that our approach could eventually be used to explain these non-typical temporal LST profiles:
\begin{itemize}
    \item \textit{Bajio region, Mexico}: the Bajio region in Mexico records a very high density of data centres managed by various providers that started operations approximately twenty years ago. 
    The stable climate, low seismic activity, and proximity to North American markets made the Bajío region a great hub for AI data centres. 
    Nevertheless, it has been recorded a serious LST increase trend (in the order of 2°C) in the last twenty years in the region, which was not identified in proximal areas~\cite{DHI_CaseStudy1}; 
    \item \textit{Aragon province, Spain}: Aragón has emerged as a major European hub for hyperscale AI data centres. 
    The region is becoming a critical node for AI, cloud computing, and, increasingly, specialized server manufacturing. 
    At the same time, the region has recorded an anomalous increase of approximately 2°C, which stands out with respect to the trends of the neighboring provinces of Spain, as well as to the global temperature increase that has been monitored in Europe~\cite{DHI_CaseStudy3};
    \item \textit{Ceara' and Piauí states, Brazil}: The north east region of Brazil identifies one of the areas in the country (together with the greater urban areas of Rio de Janeiro and Sao Paulo) with a very high concentration of data centres. 
    The surroundings of the city of Teresina, Piauí, are in particular dedicated for AI service hub. 
    At the same time, the states of Ceara' and Piauí have shown a peculiar LST increase trend centred in Teresina in the range of 2.8°C, projected to reach more than 3.5°C in the next five years, that is quite unusual with respect to other areas in the north Brazil and equatorial Brazil~\cite{DHI_CaseStudy2}.
\end{itemize}
These results become more significant when considering the population exposed to data heat island effects. 
We report in Figure~\ref{fig_TempPop} the amount of people that were resident within 10 km radius from the AI hyperscalers with respect to the LST increase that they would have experienced after the start of operations of each data centre. 
Figure~\ref{fig_TempPop} displays the result of this analysis. 
Although the object of the analysis has been focused on data centres outside of densely populated areas (as mentioned in Section~\ref{sec:methods}), it is possible to appreciate how many people (up to 343 millions in total for a LST increase up to 9°C) could be affected by the data heat island effects worldwide. 
This makes the data heat island effect a phenomenon that is very hard to consider negligible, as it may lead (like the urban heat island effect) to dramatic impact on welfare, healthcare and energy systems~\cite{IEA,UHI_2,UHI_3,UHI_7,UHI_9}. 

Given these findings, mitigation measures for data heat island effects warrant urgent consideration: a set of strategies that can be implemented in this respect are introduced in the next Section.  
%

\section{A way forward} 
\label{sec:way_forward}
Although the impact of data heat islands can be intense (as it has been previously discussed), advances in technology in the semiconductor and energy material industries, as well as methodological developments in computer science and electrical engineering, can be used to mitigate their effects. We categorise these possible strategies in two main classes, software- and hardware-based. We report in the following examples of realistic approaches that can be employed to alleviate the data heat island effect (although in most cases these strategies have not been designed to address LST increase in the context of climate transformation). Beyond technological interventions, addressing the environmental impact of AI hyperscalers also highlights the need for a new theory of intelligence that explicitly incorporates energy, information, and physical constraints. Traditional AI theories often treat intelligence as abstract computation or symbolic manipulation, ignoring the thermodynamic costs of information processing. The Matryoshka model of intelligence~\cite{cheint}, for instance, proposes a hierarchical, physically grounded framework in which cognitive, sensory, and motor processes are coupled with energy and material constraints across nested levels of complexity. By recognizing intelligence as a thermodynamically instantiated phenomenon, this framework not only provides deeper insight into natural and artificial cognition but also offers a principled foundation for designing more energy-efficient and sustainable AI systems, which could substantially reduce the heat and carbon footprint of large-scale data centers. Integrating such theory-driven principles into AI architecture and operations could complement hardware and software strategies, offering a holistic approach to mitigating the data heat island effect.
\subsection{Software-based solutions}
The strategies in this category pertain to the design, development and implementation of computational methods that would make the data processing in AI hyperscalers more efficient, hence reducing their inherent power consumption. In order to provide a complete overview of the directions that could be explored for limiting the impact of data heat island effect in this respect, we divide the strategies in this category according to the main phases for data data analysis, i.e., preprocessing, processing, and usage for decision making: 

\begin{itemize}
    \item \textit{how data are arranged}: the preprocessing operations in data analysis frameworks must not be underestimated. In the vast majority of operational scenarios, data analysis has to be performed onto records that show several nonidealities (e.g., missing, corrupted, unbalanced data, high variability in the records, low informativity of the data samples). This affects the quality of data analysis, as classic AI models would aim to reduce this uncertainty by implementing transformations (e.g., interpolation, smoothing) on the data, which could lead to artificial bias and hallucinations~\cite{UHI_9}. In this respect, considering the records as realizations of propagation and advection forces that can be used to model the properties of the data samples in the feature space can help~\cite{Marinoni_PAMI}. In fact, this choice enables the identification of the actual relevance of the data samples and features, hence supporting efficient computational procedure by removing the negligible connections across samples and focusing the algorithms’ attention on the significant and informative components of the datasets only. Core datasets can be selected from their original datasets to improve training efficiency in deep
learning, without loss of performance accuracy \cite{Weisi3}. 
    \item \textit{how data are processed}: one of the greatest challenges of classic AI algorithms is convergence to unacceptably inferior local minima during the loss minimization or training process~\cite{UHI_9, SustainableAI, OpenAI_compute, pantow, pansel}. This leads to  extreme difficulties for AI algorithms to guarantee any level of predictive accuracy without allowing AI models to use a large amount of energy, e.g., to identify all the possible solutions across the feature space~\cite{AIEnergyConsum, McKinsey2, LCA_AI, Microsoft24}. Seminal concepts such as the universality of deep, feed-forward neural networks and empirically encouraging results, although the degree of improvement in terms of power efficiency is most of the times very limited. Also, hierarchical learning based on functional learning (e.g., isogeometrical networks~\cite{GASICK2023115839}) can be used to mitigate training and convergence issues, and can increase the accuracy of the classic AI strategies whilst guaranteeing a sustainable use of the computing resources.  
    Also, pruning and compression of deep learning based models \cite{Weisi1, Weisi2} can benefit the definition and development of solution for hardware-aware AI models.
    \item \textit{how data are used}: as other anthropogenic activities, data centeres emissions can be quantified in terms of carbon footprint, i.e., the the amount of greenhouse gas emissions (gCO2) generated~\cite{UHI_1,UHI_10,UHI_2,UHI_11}. In this respect, novel strategies for innovative carbon-aware generative language model inference framework have been designed to reduce carbon footprint: the primary method for this relies on the strategic use of token generation directives while maintaining high-quality outputs~\cite{DORGEVAL2026127288,Science_masanet,SustainableAI,NPJ_water}. To achieve these goals architectures proposed in technical literature employ numerical programming for system-level optimization, balancing carbon savings with generation quality.
\end{itemize}

\subsection{Hardware-based solutions} 
In this category, we summarise some viable options to improve data centres efficiency (hence their heat release in the atmosphere) by enhancing their operational framework. As for the previous case, we list these novel approaches according to the stage of functional deployment in the AI hyperscalers, i.e., at signal processing, power management, and infrastructure level. 
\begin{itemize}
    \item \textit{how data are translated into signals}: recent progresses in integrated circuitry have enabled the development of technologies able to reduce the power consumption of AI accelerators and indeed recover the energy used for computing~\cite{BUTT2025100326, UHI_9, SustainableAI}. Adiabatic circuitry plays a key role in this respect~\cite{Pal2015}. This class of low-power integrated circuits rely on reversible logic and energy recovery, so that  reduction of energy loss  during computation can be achieved. This directly leads to reduction in heat dissipation, therefore potentially reducing the data heat island effect. Indeed, in adiabatic circuitry, power is recycled mainly through transistor switching. Although very promising, this technology show consistent limits for speed and scaling to AI models, hence requiring very complex systems (e.g., sinusoidal/trapezoidal clocks) to be implemented. 
    \item \textit{how signals are translated into power}: AI workloads are characterized by high computational intensity over long timeframes; high degree of variability, unpredictability, and nonlinear scalability of computational power usage; sensitivity to algorithmic design and implementation~\cite{AIEnergyConsum, Alberta, PowerAI, PowerHungry, McKinsey, McKinsey2}. This leads to major problems for demand and supply of AI hyperscalers within the energy grids that they are embedded in, e.g., affecting the grid stability and the accuracy of the AI models that are run onto their structures. These issues are a direct byproduct of the inability of AI hyperscalers power infrastructures to track the high variability of AI models usage. However, when power resources are set up to provide dynamic power response~\cite{Marinoni_DRS} (e.g., by intelligent battery energy support systems paired with power-load-temperature optimisation systems for resource and job allocation), the efficiency of data centres could dramatically improve, e.g., showing solid reduction of data centre downtime; protection of the infrastructure by power fluctuations, drops and spikes; and higher resilience towards irregular and unstructured AI platform usage. 
    \item \textit{how power is managed in infrastructure} The electronic components in AI hyperscalers are characterised by extremely high-power densities, often reaching magnitudes on the order of 107 W/m² \cite{Thermal1}. These extreme thermal loads, exacerbated by non-uniform workload distributions and the emergence of localized hotspots across chips, demand sophisticated thermal management strategies \cite{Thermal2}. One key approach involves multiscale coordinated cooling schemes that combine localised liquid cooling at the chip level with system-wide air cooling across the facility. This hybrid method targeted heat extraction from critical hotspots while maintaining overall thermal balance throughout the infrastructure \cite{Thermal3}. In addition to active cooling methods, passive cooling technologies offer complementary benefits by alleviating structural thermal load without increasing operational energy demand. Among these, passive radiative cooling has garnered particular interest. This technique engineers the optical properties of outdoor-exposed surfaces, such as building envelopes, by incorporating high solar reflectivity to suppress solar heat gain and strong thermal emissivity in long-wave infrared spectrum to enable radiative heat dissipation, thereby reducing the overall thermal load on infrastructure \cite{Thermal4,Thermal5,Thermal6}. Recent advancements have translated this passive cooling technology into practical applications. 
    In particular,  passive radiative cooling coating based on polymer–nanoparticle composite has been applied across various real-world scenarios, including residential structures, urban infrastructure, and agricultural storage facilities \cite{Thermal7}. Reported results indicate cooling load reductions ranging from 10$\%$ to 40\% after applying the coating, contingent on site-specific factors such as surface albedo, orientation, and geographic climate conditions. In the context of AI hyperscalers, where thermal loads are not only intense, but spatially heterogeneous, such passive approaches can play a structural role in limiting the baseline thermal burden imposed on active cooling systems.
\end{itemize}

\section{Conclusions}
\label{sec:conclusions}

The increasing demand for AI-based services, processes and operations led to the proliferation of data centres worldwide that are extremely power hungry. In this paper, we provide the first assessment of the environmental impact of AI hyperscalers. We focus our attention on the heat dissipation of data centres, which is directly connected to the energy consumption required for the operations of the AI hyperscalers. We investigate, by means of a multimodal multiscale architecture, the land surface temperature change occurred as a consequence of the start of operations of the data centres. Our analysis spans over the last decades (from 2004 to 2024), taking advantage of the plethora of remotely sensed temperature measurements  acquired by satellites worldwide. 
Our study shows a non negligible and rather remarkable impact of the AI data centres on their local regions, which is consistent across the data centres worldwide and extends for several kilometers around the AI hyperscalers. 
The consistency, scale and extent of these effects lead to think that the creation of local climate zones induced by data centres - that we call the \textit{data heat island effect} - is real and significant, especially in the context of global warming and climate transformation. 

Consequently, the data heat island effect could affect the welfare, healthcare, energy, and  demographic systems. 
Since the trends of data centre energy consumption are expected to show a steep growth in the foreseeable future, the data heat island effect could solidly  become an additional factor for environmental and industrial sustainability in the changing climate, hence having a robust impact on communities at local, regional, and international level, thus demanding to be studied in complex multi-hazard systems. 
To this aim, in this paper we provide an overview of potential solutions to alleviate the data heat island effect, which could be further expanded into mitigation policies for future climate and socioeconomic scenarios.

\bibliography{sn-bibliography}

\end{document}